\documentclass[epj]{svjour}

\usepackage{epsfig}
\usepackage{wrapfig}                                                  
          
\newcommand{\pom}{{I\!\!P}}
\newcommand{\xpom}{x_\pom}
\newcommand{\av}[1]{\mbox{$ \langle #1 \rangle $}}

\def\Journal#1#2#3#4{{#1} {\bf #2}, (#3) #4}

\def\NPB{{\em Nucl. Phys.}   {\bf B}}
\def\PLB{{\em Phys. Lett.}   {\bf B}}
\def\PRL{\em Phys. Rev. Lett.}
\def\PRD{{\em Phys. Rev.}    {\bf D}}
\def\ZPC{{\em Z. Phys.}      {\bf C}}
\def\EJC{{\em Eur. Phys. J.} {\bf C}}

\newcommand{\etal}{{\em et al.}}

\begin{document}

\title{Diffractive Final States with the H1 Detector at HERA}

\author{Frank-Peter Schilling\inst{} (for the H1 collaboration)
\thanks{e-mail: {\tt fpschill@mail.desy.de}}
\thanks{Talk presented at the Intl. Europhysics Conference on High Energy
Physics EPS 2003, Aachen, July 2003}
}

\institute{DESY, Notkestr. 85, D-22603 Hamburg, Germany}

\date{Received: date / Revised version: date}

\abstract{ Recent measurements of diffractive dijet and charm quark
  production in electron-proton collisions using the H1 detector at
  HERA are presented, where the exchanged photon is either almost real
  or highly virtual.  The data are compared with leading and
  next-to-leading order QCD calculations based on the diffractive
  parton distributions obtained from a recent DGLAP QCD analysis of H1
  inclusive diffractive deep-inelastic scattering data, thus testing
  QCD factorization in diffractive $ep$ interactions.
}

\maketitle


\section{Introduction}
\label{intro}

Understanding diffraction in hadronic interactions at high energies,
where at least one of the beam hadrons remains intact, losing only a
small fraction $x_\pom$ of its incident longitudinal momentum,
represents one of the most important challenges in Quantum
Chromodynamics (QCD). The $ep$ collider HERA offers the unique
possibility to study hard diffractive processes, such as dijet and
heavy quark production, over a wide range of photon virtualities
$Q^2$.

Recent high precision measurements of the inclusive diffractive
deep-inelastic scattering (DIS) cross section by the H1 collaboration
\cite{f2d97} have been used to extract diffractive parton
distributions (dpdf's) of the proton by means of a DGLAP QCD fit. If
QCD factorization is valid in diffractive $ep$ scattering, these
dpdf's are universal and can be used to predict cross sections for
exclusive hard diffractive processes such as jet and heavy quark
production.

The proof of QCD factorization in diffractive $ep$ interactions
\cite{collins} is restricted to deep-inelastic scattering at large
$Q^2$ and not valid for the case of an almost real photon
(photoproduction).  Furthermore, a severe breakdown of factorization
has been observed when using dpdf's obtained at HERA to predict
diffractive jet production in hadron-hadron interactions at the
Tevatron \cite{cdf}.

Here, QCD factorization is tested in DIS to next-to-leading order
(NLO) by comparing cross section measurements for dijet and $D^*$
meson production in diffractive DIS with predictions obtained from
convoluting the dpdf's from \cite{f2d97} with NLO order matrix
elements for dijet and charm quark production \cite{nlopaper}.  In
addition, diffractive dijet photoproduction data are compared with
leading order (LO) Monte Carlo predictions \cite{gpjets}.

\subsection{Kinematics}

\begin{wrapfigure}[16]{L}{0.5\linewidth}
\centering
\epsfig{file=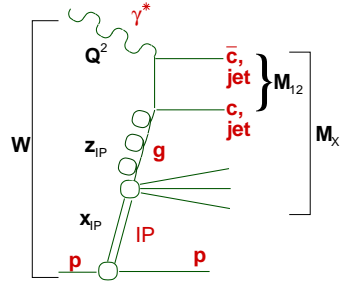,width=0.99\linewidth}
\caption{Diagram for diffractive jet or charm production at HERA.}
\label{fig:feyn}
\end{wrapfigure}

Fig.~\ref{fig:feyn} shows an example diagram for diffractive dijet or
charm quark production at HERA.

The photon (virtuality $Q^2$) emitted from the beam electron (not
shown) interacts with the proton, which loses only a small fraction
$x_\pom$ of its incident momentum and stays intact. The longitudinal
momentum fraction of the parton entering the hard scattering process
relative to the diffractive exchange is labelled $z_\pom$.  A pair of
high transverse momentum ($p_T$) jets or heavy quarks is produced. The
photon-proton centre-of-mass energy is labelled $W$ and the invariant
masses of the diffractively produced system and of the two partons
emerging from the hard sub-process are denoted $M_X$ and $M_{12}$,
respectively.  The inelasticity variable $y$ is given by $ys=W^2+Q^2$,
where $s$ is the squared $ep$ centre-of-mass energy.

\subsection{H1 Diffractive Parton Distributions}

QCD factorization permits to express the diffractive DIS $\gamma^*p$
cross section as a convolution of diffractive parton distributions
$p_i^D$ and partonic cross sections $\hat{\sigma}^{\gamma^*i}$:
\begin{equation}
\frac{{\rm d^2} \sigma_{\gamma^*p}^D}
{{\rm d} x_\pom \ {\rm d} t} \ \sim \
\sum_i \
\hat{\sigma}^{\gamma^*i}(x,Q^2) \otimes
p_i^D(x,Q^2,x_\pom,t) \ ,
\label{equ:diffpdf}
\end{equation}
where $t$ is the squared 4-momentum transferred at the proton vertex.
The dpdf's are universal and obey the DGLAP evolution equations.  In
the QCD fit to the H1 inclusive diffractive DIS data in \cite{f2d97},
{\em Regge factorization} was assumed as well (consistent with the
data within the present level of precision), i.e.
\begin{equation}
p_i^D(x,Q^2,x_\pom,t) = f_\pom(x_\pom,t) \ p_i^\pom(z=x/\xpom,Q^2) \ ,
\end{equation}
so that the shape of the dpdf's is independent of $x_\pom,t$ and the
normalization is controlled by a flux factor $f_\pom$, for which a
value for the {\em pomeron intercept} $\alpha_\pom(0)=1.173$ is used.
The parton distributions $p_i^\pom$, as determined from both the LO
and NLO QCD fits, are dominated by the gluon density and extend to
large fractional momenta $z$.

\section{Dijet Production in Diffractive DIS}

To test QCD factorization for diffractive dijet production in DIS, the
H1 cross sections from \cite{h1jets} were used, corresponding to the
kinematic range $4<Q^2<80 \rm\ GeV^2$ and $x_\pom<0.05$. The cross
sections were corrected to asymmetric cuts on the jet transverse
momentum $p_{T,1(2)}>5(4) \rm\ GeV$, to facilitate comparisons with
NLO calculations.

\begin{wrapfigure}[23]{L}{0.5\linewidth}
\centering
\epsfig{file=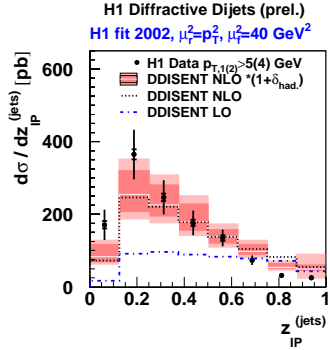,width=0.99\linewidth}
\caption{Diffractive DIS dijet cross section as a function of 
$z_\pom^{(jets)}$, an estimator for the parton momentum fraction
of the diffractive exchange entering the hard subprocess.}
\label{fig:dis1}
\end{wrapfigure}

The NLO dpdf's obtained in \cite{f2d97} were convoluted with NLO
matrix elements for dijet production by means of the DISENT
\cite{disent} program, as suggested in \cite{hautmann}.  The
renormalization and factorization scales were set to the average $p_T$
of the two highest $p_T$ partons.  $\alpha_s$ was set via
$\Lambda_{QCD}^{n=4}=200 \rm\ MeV$, as in the QCD fit.  To allow
comparisons with the measured data, the same jet algorithm as for the
data, as well as hadronization corrections, were applied.

\begin{figure}
\centering
\epsfig{file=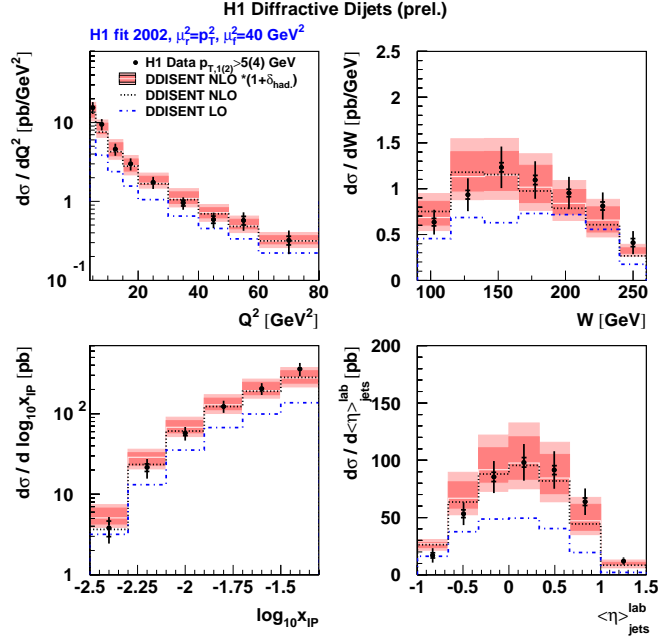,width=0.99\linewidth}
\caption{
  Diffractive DIS dijet cross sections differential in $Q^2$, $W$,
  $\log_{10} x_\pom$ and the average dijet pseudorapidity
  $\av{\eta}_{jets}^{lab}$.  }
\label{fig:dis2}
\end{figure}

Comparisons of the LO and NLO QCD calculations with the dijet data are
shown in Figs.~\ref{fig:dis1} and \ref{fig:dis2}.  The size of the NLO
corrections is on average more than a factor 2 (increasing with
decreasing $p_T$ and $Q^2$).  The inner error band of the NLO
calculations represents the renormalization scale uncertainty, whereas
the outer band includes the uncertainty in the hadronization
corrections.  Within the uncertainties, the data are well described in
both shape and normalization by the NLO calculations based on the
dpdf's from \cite{f2d97}, in agreement with QCD factorization.

\section{{\boldmath$D^*$} Meson Production in Diffractive DIS}

In \cite{h1dstar}, cross sections for $D^*$ meson production in
diffractive DIS were published for the kinematic range $2<Q^2<100 \rm\ 
GeV^2$, $x_\pom<0.04$ and $p^*_{T,D^*}> 2 \rm\ GeV$, where the latter
variable corresponds to the transverse momentum of the $D^*$ meson in
the photon-proton centre-of-mass frame.

\begin{figure}
\centering
\epsfig{file=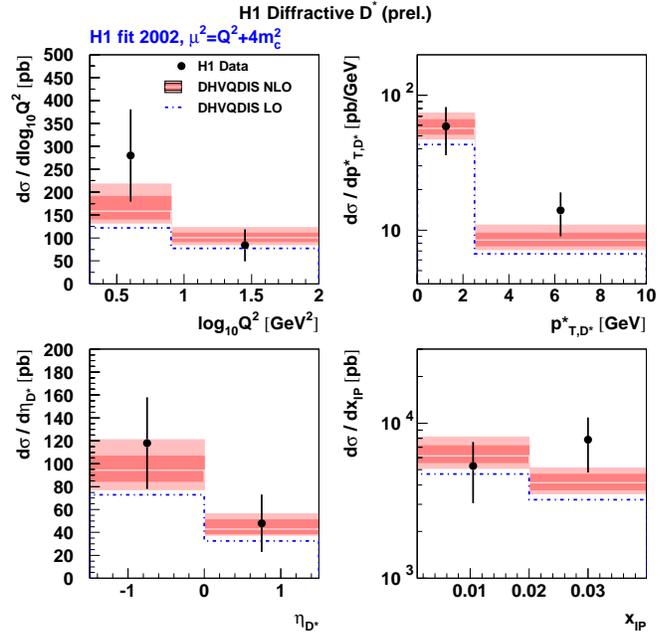,width=0.99\linewidth}
\caption{Diffractive DIS $D^*$ meson cross sections differential
in $Q^2$, $p^*_{T,D^*}$, $\eta_{D^*}$ and $x_\pom$.
}
\label{fig:dstar}
\end{figure}

Both LO and NLO QCD calculations were performed using the diffractive
version \cite{dhvq} of HVQDIS \cite{hvqdis}, interfaced to the H1
dpdf's.  The renormalization and factorization scales were set to
$\mu^2=Q^2+4m_c^2$.  Further parameter values are a charm quark mass
of $m_c=1.5 \rm\ GeV$, a $c\rightarrow D^*$ hadronization fraction of
$f(c\rightarrow D^*)=0.233$ and $\epsilon=0.078$ for the used Peterson
fragmentation function.

A comparison of the calculations with the $D^*$ data is shown in
Fig.~\ref{fig:dstar}. The inner error band of the NLO calculation
represents the renormalization scale uncertainty, whereas the outer
error band includes variations of $m_c$ and $\epsilon$. Within the
uncertainties, the data are well described in both shape and
normalization by the NLO calculations, using the dpdf's from
\cite{f2d97}, supporting the idea of QCD factorization.

\section{Dijets in Diffractive Photoproduction}

In \cite{gpjets}, H1 has presented a measurement of diffractive dijet
photoproduction, i.e. for $Q^2\sim0$.  The cross sections correspond
to the kinematic range $Q^2<0.01 \rm\ GeV^2$, $0.3<y<0.65$,
$x_\pom<0.03$ and $p_{T,1(2)}>5(4) \rm\ GeV$, where jets are defined
using the $k_T$ algorithm.

\begin{figure}
\centering
\epsfig{file=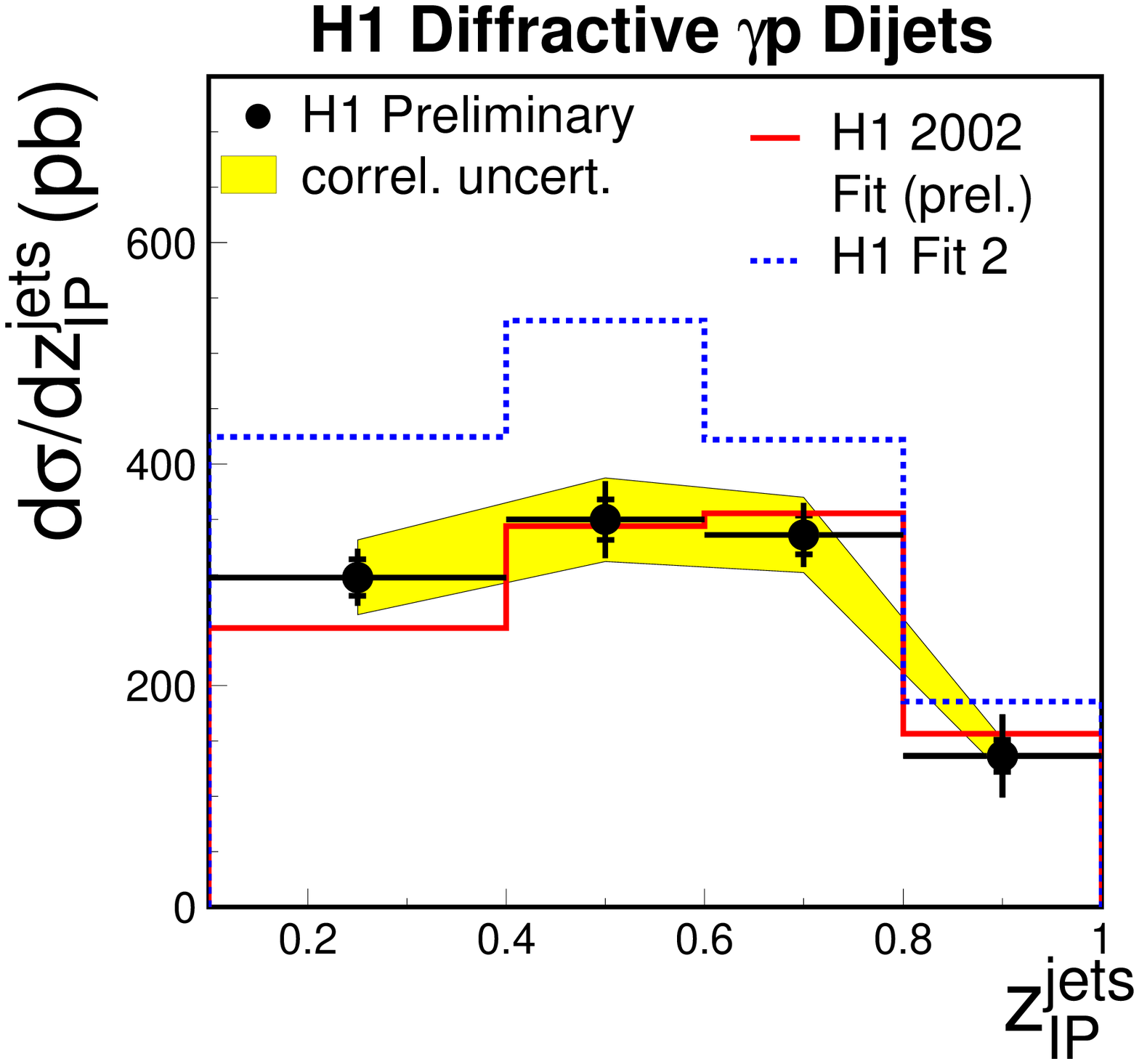,width=0.46\linewidth}
\epsfig{file=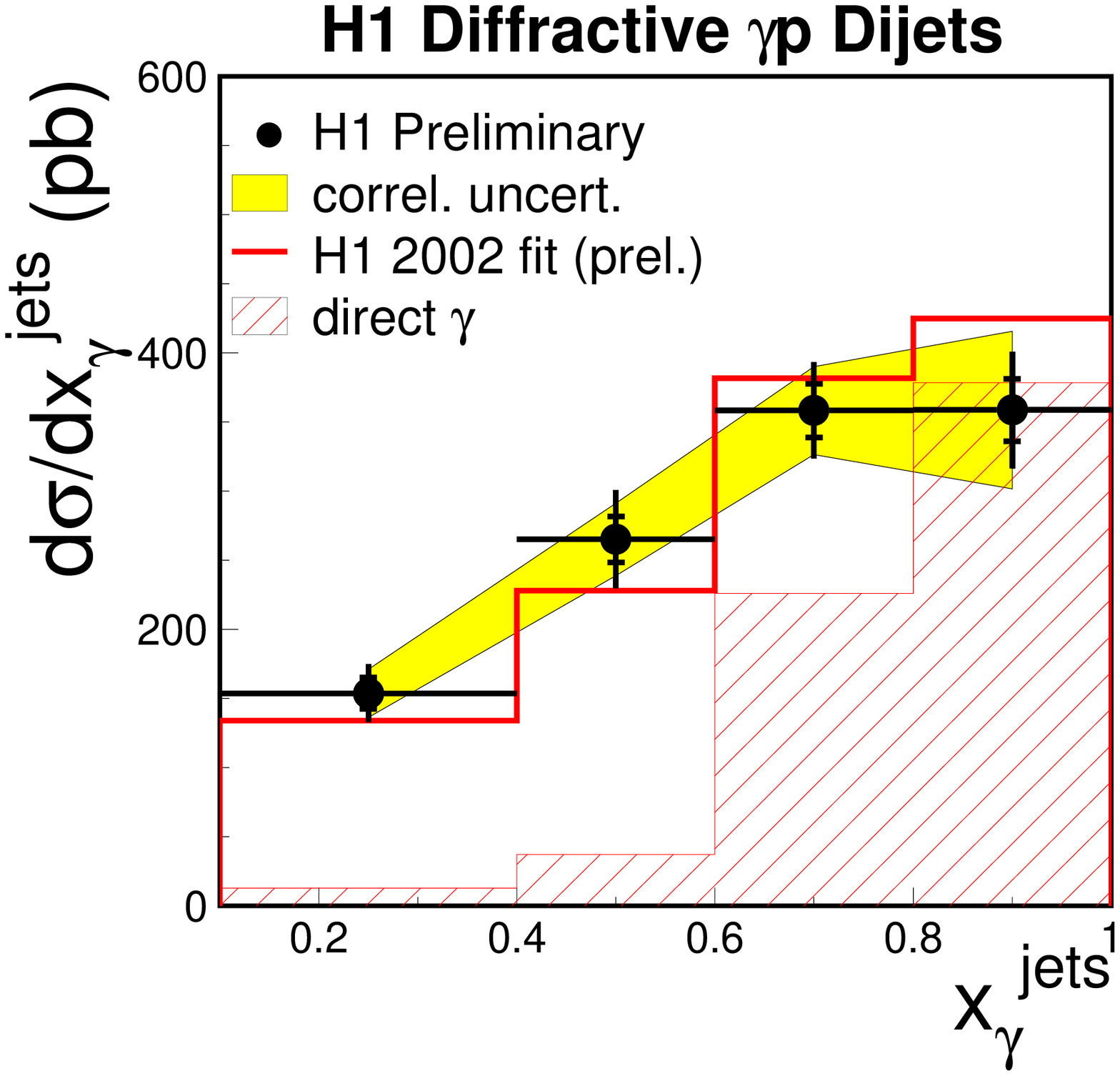,width=0.46\linewidth}
\caption{Diffractive dijet photoproduction cross sections
differential in $z_\pom^{jets}$ and $x_\gamma^{jets}$.
}
\label{fig:gp1}
\end{figure}

\begin{figure}
\centering
\epsfig{file=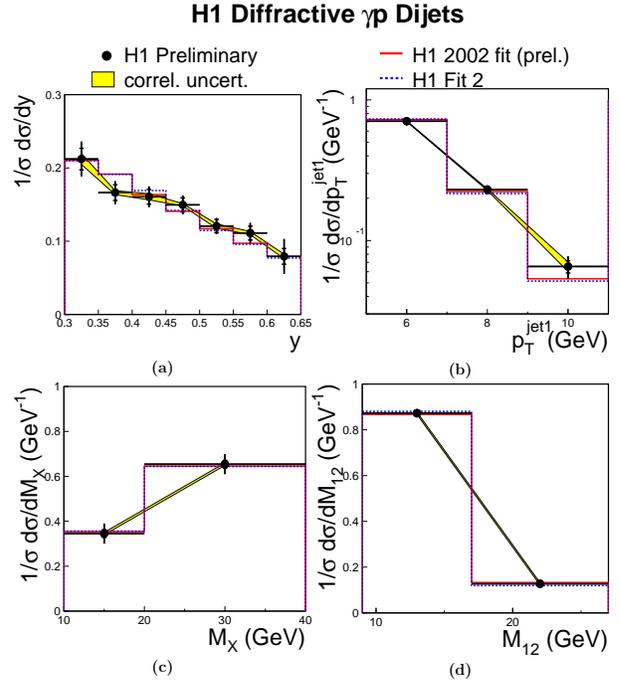,width=0.9\linewidth}
\caption{Normalized diffractive dijet photoproduction cross sections
differential in $y$, $p^{jet \ 1}_T$, $M_X$ and $M_{12}$.
}
\label{fig:gp2}
\end{figure}

The data are compared in Figs.~\ref{fig:gp1} (absolute cross sections)
and \ref{fig:gp2} (normalized cross sections) with LO Monte Carlo
predictions using both the LO version of the dpdf's from \cite{f2d97},
as well as an earlier version from \cite{h1f2d94}.  $x_\gamma^{jets}$
is an estimator for the momentum fraction of the photon entering the
hard process, which is used to distinguish between direct and resolved
photon interactions, where in the latter case the photon develops
hadronic structure, and a parton from the photon with momentum
fraction $x_\gamma$ enters the hard process.

For both direct and resolved interactions, the dijet data are well
described by the Monte Carlo predictions if the new H1 dpdf's from
\cite{f2d97} are used. Thus, the data are not consistent with a
breakdown of factorization even in resolved photoproduction, which
resembles hadron-hadron scattering where the presence of the second
hadron may lead to a suppression of the rate of diffractive events due
to spectator interactions.

\section{Conclusions}

The new H1 diffractive parton distributions have been interfaced with
NLO QCD calculations and compared with dijet and $D^*$ production data
in diffractive DIS.  The good agreement within the uncertainties lends
support for the validity of QCD factorization in diffractive DIS, now
tested to next-to-leading order.  In addition, also photoproduction
diffractive dijet measurements are described in shape and
normalization by Monte Carlo predictions using these dpdf's, not
indicating any breakdown of factorization in diffractive $ep$
interactions at HERA.



\end{document}